\documentclass[fleqn,usenatbib]{mnras}

\usepackage{graphicx}	
\usepackage{amsmath}	
\usepackage{amssymb}	
\usepackage{bm}	        
\usepackage{mathtools}
\usepackage[utf8]{inputenc}
\usepackage[english]{babel}
\usepackage[section] {placeins}
\usepackage{textcomp,gensymb}
\usepackage{color}
\usepackage{xcolor}
\usepackage{multirow}
\usepackage{makeidx}
\usepackage{xspace}
\usepackage{float}
\usepackage{upgreek} 
\usepackage{comment}                                                                               
\usepackage{verbatim}
\usepackage{subcaption}
\usepackage{etoolbox}
\usepackage[normalem]{ulem} 

\usepackage{media9}
\usepackage{tikz}
\newif\ifthreed
\threedtrue 

\definecolor{ForestGreen}{HTML}{2e8b21}

\usepackage{scalerel}
\usepackage{tikz}
\usetikzlibrary{svg.path}

\definecolor{orcidlogocol}{HTML}{A6CE39}
\tikzset{
  orcidlogo/.pic={
    \fill[orcidlogocol] svg{M256,128c0,70.7-57.3,128-128,128C57.3,256,0,198.7,0,128C0,57.3,57.3,0,128,0C198.7,0,256,57.3,256,128z};
    \fill[white] svg{M86.3,186.2H70.9V79.1h15.4v48.4V186.2z}
                 svg{M108.9,79.1h41.6c39.6,0,57,28.3,57,53.6c0,27.5-21.5,53.6-56.8,53.6h-41.8V79.1z M124.3,172.4h24.5c34.9,0,42.9-26.5,42.9-39.7c0-21.5-13.7-39.7-43.7-39.7h-23.7V172.4z}
                 svg{M88.7,56.8c0,5.5-4.5,10.1-10.1,10.1c-5.6,0-10.1-4.6-10.1-10.1c0-5.6,4.5-10.1,10.1-10.1C84.2,46.7,88.7,51.3,88.7,56.8z};
  }
}

\newcommand\mstar{\ensuremath{M_*}\xspace}
\newcommand\mhavg{\ensuremath{\langle [\rm{M/H}] \rangle}\xspace}

\newcommand\orcidicon[1]{\href{https://orcid.org/#1}{\mbox{\scalerel*{
\begin{tikzpicture}[yscale=-1.,transform shape]
\pic{orcidlogo};
\end{tikzpicture}
}{|}}}}

\usepackage{hyperref} 

\usepackage{etoolbox}
\makeatletter
\newcommand\sendemail[3]{
\edef\@tempa{mailto:#1?subject=#2 }%
\edef\@tempb{\expandafter\html@spaces\@tempa\@empty}%
\href{\@tempb}{#3}}

\catcode\%=11
\def\html@spaces#1 #2{#1
\catcode\%=14
\makeatother

\newcommand{\orcid}[2]{\href{http://orcid.org/#2}{#1}}

\title[The Stellar FMR]{The stellar Fundamental Metallicity Relation: the correlation between stellar mass, star-formation rate and stellar metallicity}

\author[\orcid{T. J. Looser}{0000-0002-3642-2446}~et al.]{\parbox{\textwidth}{
\orcid{Tobias J. Looser}{0000-0002-3642-2446}$^{1,2}$\thanks{E-mail: tjl54@cam.ac.uk}
\orcid{Francesco D'Eugenio}{0000-0003-2388-8172}$^{1,2}$,
\orcid{Joanna M. Piotrowska}{0000-0003-1661-2338}$^{1,2,3}$,
\orcid{Francesco Belfiore}{0000-0002-2545-5752}$^{4}$,
\orcid{Roberto Maiolino}{0000-0002-4985-3819}$^{1,2,5}$,
\orcid{Michele Cappellari}{0000-0002-1283-8420}$^6$,
\orcid{William M. Baker}{0000-0003-0215-1104}$^{1,2}$ and
\orcid{Sandro Tacchella}{0000-0002-8224-4505}$^{1,2}$
}
\vspace{0.4cm}
\\
\parbox{\textwidth}{
$^{1}$Kavli Institute for Cosmology, University of Cambridge, Madingley Road, Cambridge, CB3 0HA, UK\\
$^{2}$Cavendish Laboratory - Astrophysics Group, University of Cambridge, 19 JJ Thomson Avenue, Cambridge, CB3 0HE, UK\\
$^{3}$Cahill Center for Astronomy and Astrophysics, California Institute of Technology, Pasadena, CA 91125, USA\\
$^{4}$INAF -- Osservatorio Astrofisico di Arcetri, Largo E. Fermi 5, I-50125, Florence, Italy\\
$^{5}$Department of Physics and Astronomy, University College London, Gower Street, London WC1E 6BT, UK\\
$^{6}$Sub-department of Astrophysics, Department of Physics, University of Oxford, Denys Wilkinson Building, Keble Road, Oxford OX1 3RH, UK
}
}

\date{Accepted n/a. Received n/a; in original form 2023 May 21}

\pubyear{2023}

\begin{document}
\label{firstpage}
\pagerange{\pageref{firstpage}--\pageref{lastpage}}
\maketitle

\begin{abstract}
We present observational evidence for a stellar Fundamental Metallicity Relation (FMR), a smooth relation between stellar mass, star-formation rate (SFR) and the light-weighted stellar metallicity of galaxies (analogous to the well-established gas-phase FMR).
We use the flexible, non-parametric software pPXF to reconstruct simultaneously the star-formation and chemical-enrichment history of a representative sample of galaxies from the local MaNGA survey. We find that
(i) the metallicity of individual galaxies increases with cosmic time and (ii) at all stellar masses, the metallicity of galaxies is progressively higher, moving from the star-burst region above the main sequence (MS) towards the passive galaxies below the MS, manifesting the stellar FMR. 
These findings are in qualitative agreement with theoretical expectations from IllustrisTNG, where we find a mass-weighted stellar FMR.
The scatter is reduced when replacing the stellar mass $M_{*}$ with $M_{*}/R_{\rm e}$ (with $R_{\rm e}$ being the effective radius), in agreement with previous results using the velocity dispersion $\sigma_{\rm e}$, which correlates with $M_{*}/R_{\rm e}$. Our results point to starvation as the main physical process through which galaxies quench, showing that metal-poor gas accretion from the intergalactic/circumgalactic medium -- or the lack thereof -- plays an important role in galaxy evolution by simultaneously shaping both their star-formation and their metallicity evolutions, while outflows play a subordinate role. This interpretation is further supported by the additional finding of a young stellar FMR, tracing only the stellar populations formed in the last 300 Myr. This suggests a tight co-evolution of the chemical composition of both the gaseous interstellar medium and the stellar populations, where the gas-phase FMR is continuously imprinted onto the stars over cosmic times. 
\end{abstract}

\begin{keywords}
galaxies: formation — galaxies: evolution — galaxies: star formation — galaxies: stellar content — galaxies: statistics
\end{keywords}

\section{Introduction} \label{sec:intro}
The intertwined co-evolution of the chemical composition of both the gaseous interstellar medium (ISM) and the stellar populations within a galaxy is governed by numerous physical processes. These processes include star formation via gravitational collapse of baryonic clouds in the ISM \citep[see e.g.][for a review]{Kennicutt2012ARA&A..50..531K}, nucleosynthesis of metals in stellar cores, supernovae explosions and stellar winds returning metals to the ISM \citep[e.g.][]{Maiolino_Mannucci_2019}, (pristine) gas accretion from the intergalactic/circumgalactic medium (IGM/CGM), or gas outflows caused by stellar winds, shocks, feedback from active galactic nuclei (AGN) and supernovae feedback \citep[e.g.][]{Veilleux2005ARA&A..43..769V}. 

The metal content of stellar populations and ISM is therefore highly informative for galaxy evolution studies, because metallicity (defined as the fractional mass of metals relative to the mass of all baryons) carries the imprint of the physical processes driving the baryon cycle over cosmic times \citetext{see \citealp{Maiolino_Mannucci_2019} for a review}. This sensitive interplay of physical processes leads naturally to a~set of observable scaling relations between galaxy metallicity and other global galaxy properties.


Previous studies have shown that both the stellar and gas metallicities scale with stellar mass $M_*$ \citep[e.g.][]{Tremonti2004, Gallazzi2005}, known as mass--metallicity relations (MZR), stating that more massive galaxies are on average more chemically enriched than lower mass galaxies. The gas MZR has been shown to hold to at least $z \sim 3.3$ \citep{Sanders2021ApJ...914...19S, Li2022arXiv221101382L}, with growing evidence that it holds up to $z \sim 10$ \citep{Curti2023,Nakajima2023arXiv230112825N}. These findings together suggest that the stellar MZR reflects the imprint of the observed gas MZR on the stellar populations by the baryon cycle over cosmic times. 

In addition to its primary dependence on $M_*$, the gas metallicity has been shown to depend on the recent star-formation activity of a galaxy (on the timescale of \mbox{$\sim 10$ Myr}), a relation known in the literature as the gas-phase fundamental metallicity relation \citep[henceforth: gFMR; e.g.,][]{Ellison_2008, Mannucci2010, Curti2020, Baker_rFMR_10.1093/mnras/stac3594}. These studies use either SFR, the specific SFR (sSFR=SFR/$M_*$), or the distance to the star-forming main sequence ($\Delta_{MS}$) as a quantitative metric for the recent star-formation activity in galaxies, see \citet{Salim2014ApJ...797..126S} for a discussion.

While the gFMR has been studied in detail in the literature, it is to date unclear how and even \emph{if} the gFMR leaves its imprint on the \emph{stellar} metallicity (see e.g. \cite{Pipino2014MNRAS.441.1444P}, based on the gas-regulator model by \cite{2013ApJ...772..119L}, for a discussion).

\citet{Peng2015} present distinct stellar MZRs for a binary classification into star-forming (SF) and quiescent galaxies in the Sloan Digital Sky Survey (SDSS), showing that SF galaxies exhibit, on average, lower metallicities than their quiescent counterparts at fixed stellar mass. This trend is present across the whole stellar mass range, with the difference becoming more significant at the low end of the galaxy mass function. \citet{Trussler2020} add `green-valley' galaxies as a third classifier, and show that this population follows a MZR with a normalization between SF and passive galaxies. Similar results are also found in large hydrodynamical cosmological simulations, e.g. in EAGLE \citep{Schaye2015}, where \citet{De_Rossi_2018} find equivalent trends in MZR by splitting their sample into three bins of sSFR.

\cite{Scott2017MNRAS.472.2833S}, \cite{
Li2018MNRAS.476.1765L} and \cite{Neumann2021MNRAS.508.4844N} further show that spiral galaxies with typically younger stellar populations have lower stellar metallicities compared to older early-type galaxies at all stellar masses. 

\citet{Lu2023} used data from the Mapping Nearby Galaxies at Apache Point Observatory (MaNGA) survey \citep{Bundy2015} to show that galaxy stellar metallicity exhibits a continuous correlation with stellar age at $z \lesssim 0.1$, a trend that \citet{Cappellari2023} showed to be already in place at $z \approx 0.8$, using data from the LEGA-C survey \citep{vanderwel+2022}.
	
In this paper we complement these previous studies by studying MZR as a function of \emph{recent} star-formation activity. Instead of focusing on stellar ages, which trace and average past star-formation activity of galaxies over long timescales, we study the stellar MZR as a \emph{continuous} function of SFR, similarly to studies of the gMZR, which established the existence of a gFMR. 

More concretely, in this work we analyse stellar metallicities, recovered using an astro-archaeological approach, as a function of two parameters: (i) stellar mass ($M_{*}$); and (ii) distance of a galaxy from the star-forming main sequence ($\Delta_{MS}$) -- as our quantitative measure of the star-forming activity of a galaxy -- in a statistically significant number of observed galaxies. We propose a `stellar Fundamental Metallicity Relation' (sFMR) analogous to the successful gFMR. Further, we present observational evidence that the sFMR not only holds for the cumulative stellar metallicity of all stellar populations (i.e. averaged over all cosmic times), but also when considering only the youngest stellar populations, which were most recently formed in galaxies.

The layout of this work is as follows. In Section \ref{sec:Data} we describe the data and the techniques this study is based on. In Section \ref{sec:FMR} we present the sFMR in MaNGA, the sFMR in IllustrisTNG and the FMR of the young stellar populations. In Section \ref{sec:Discussion} we discuss our results and present a literature comparison. In Section \ref{sec:Summary} we summarize the main findings and conclusions of this work.

Throughout this work, we assume a Salpeter \citep[][]{Salpeter1955} initial mass function (IMF) and a $\Lambda \textrm{CDM}$ cosmology with the following parameters: $H_0 = 70$ km $\textrm{s}^{-1}$/ Mpc, $ \Omega_{\textrm{M}} = 0.3$ and $\Omega_\Lambda = 0.7$.

\section{Data and Data processing} \label{sec:Data}
\subsection{The MaNGA sample}
This work is based on spatially resolved spectroscopic data from the Mapping Nearby Galaxies at Apache Point Observatory (MaNGA) survey \citep{Bundy2015}. 
The MaNGA survey is one of the core programs of the fourth-generation Sloan Digital Sky Survey (SDSS-IV), providing spatially resolved spectroscopic data for a sample of ca. 10,000 galaxies in the local Universe with redshifts $z \lesssim0.15$. For this work, we use DR17, the final data release of SDSS-IV \citep{Abdurro2022ApJS..259...35A}.

\begin{figure}
\includegraphics[width=1.0\columnwidth]{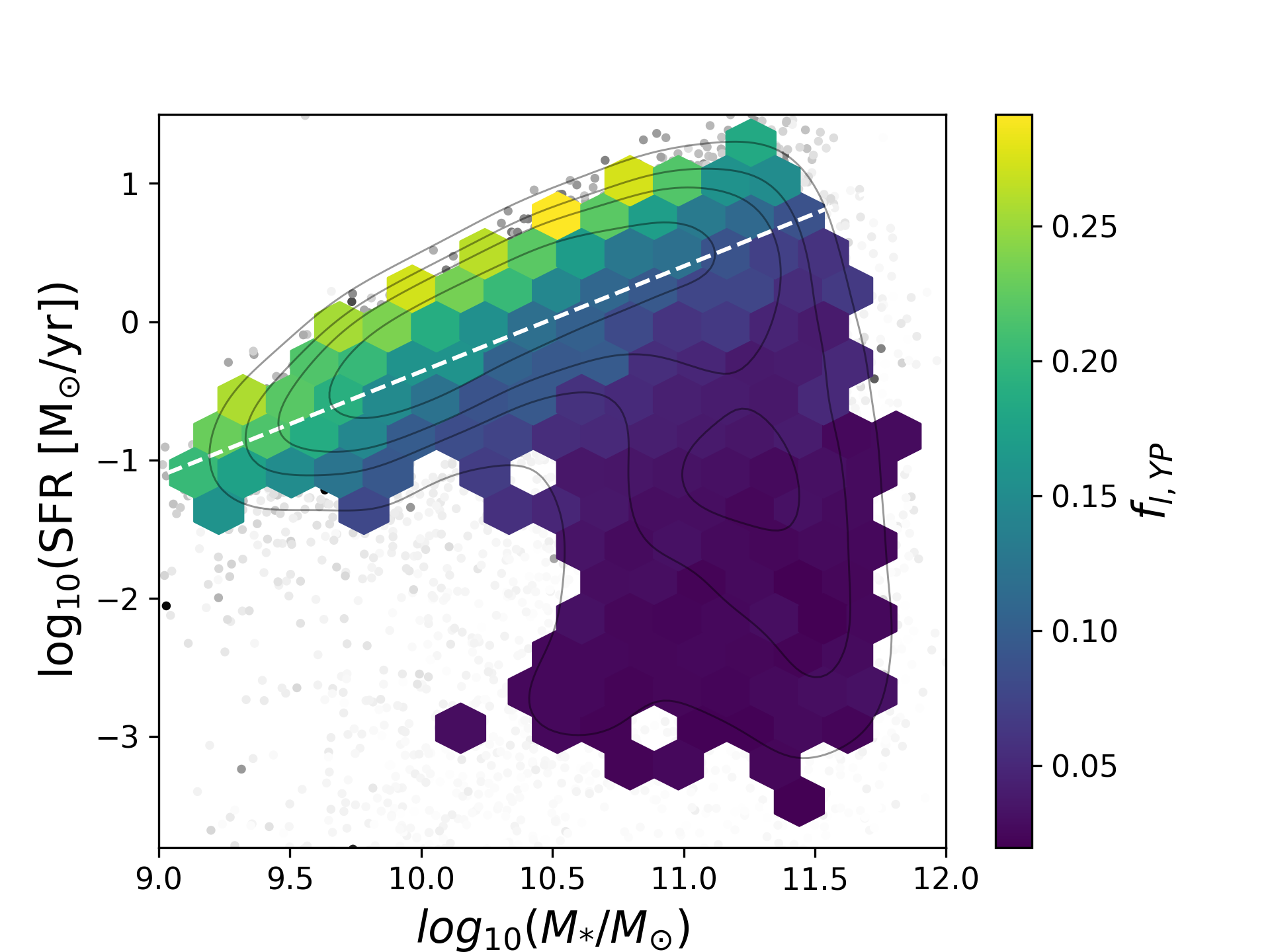}
\caption{$M_\ast-\rm{SFR}$ diagram of our MaNGA galaxy sample, color-coded by the fraction of light originating from stellar population younger than $10^{8.5} $\ yr, $f_{l,YP}$. Each hexagonal bin comprises at least 15 galaxies. The white line represents our best fit of the star-forming Main Sequence (MS) in MaNGA.
\label{fig:f_yp}}
\end{figure}

MaNGA observed with 17 fiber-bundle integral field units (IFUs) that vary in fibre number from 19 fibres to 127. Two dual-channel spectrographs provide a wavelength coverage over 3600--10300 Å at a resolution of R $\sim$ 2000. MaNGA targets are selected using NSA redshifts and $i$-band luminosity to have a total stellar mass of $ M_* \gtrsim 10^{9.5} M_{\odot}$, to achieve a uniform radial coverage in terms of the effective radius ($R_{\rm e}$, where the sample covers galaxies to 1.5 and 2.5 $R_{\rm e}$ for the primary and secondary samples, respectively) and to approximately achieve a flat distribution in stellar mass. For more details on the MaNGA survey we would like to refer to the existing literature. Specifically to \citet{Drory2015AJ....149...77D} for a description of the instrument, to \citet{Wake2017AJ....154...86W,Law2015AJ....150...19L} for the target selection and observing strategy and to \citet{Law2016,Yan2016AJ....151....8Y,Law2021AJ....161...52L} for the data reduction.

In this work, we leverage the high quality of MaNGA’s spatially resolved spectra, to study stellar populations in galaxies  at an unprecedented combination of high signal-to-noise ratio ($S/N$) and large sample size. This unique data set allows us to investigate physical processes within nearby galaxies by combining astro-archaeological
and statistical approaches across a wide range of stellar masses, morphologies and other distinctive galaxy characteristics.

\subsection{IllustrisTNG}
In order to compare our MaNGA results with state-of-the-art cosmological simulations, 
we collect an analogous sample of simulated galaxies from the $(100\ \rm cMpc)^3$ volume run of the 
IllustrisTNG simulation suite \citep{Marinacci2018, Naiman2018, Nelson2018, Pillepich2018b, Springel2018}. 
More precisely, we use the publicly available redshift $z=0$ TNG100-1 Subfind catalogue \citep{Nelson2019}
to extract stellar masses and mass-weighted mean metal mass fractions of star particles
(mass-weighted stellar metallicities). To ensure a consistent comparison with the MaNGA field of view, all 
quantities are measured within twice the stellar half-mass radius for a given subhalo (i.e.~galaxy).
We further make use of catalogues provided by \cite{Donnari2019} and \cite{Pillepich2019}, to extract
SFR values averaged over 10~Myr of the simulation to match the timescale associated with H$\upalpha$-based 
estimate of SFR in observations. We also convert $M_\ast$ and SFR from Chabrier
to Salpeter IMF by dividing the catalogue values by constant factors of 0.61 and 0.63 respectively
\citep{Madau2014}. When normalising metallicity by solar values we use the solar metallicity, defined as the bulk mass fraction
of metals, estimated by \cite{Asplund2009} of \mbox{$\rm Z_\odot$ = 0.0142}, to ensure consistency with spectral fitting performed in MaNGA. 

\begin{figure}
\centering
\includegraphics[width=.77\columnwidth]{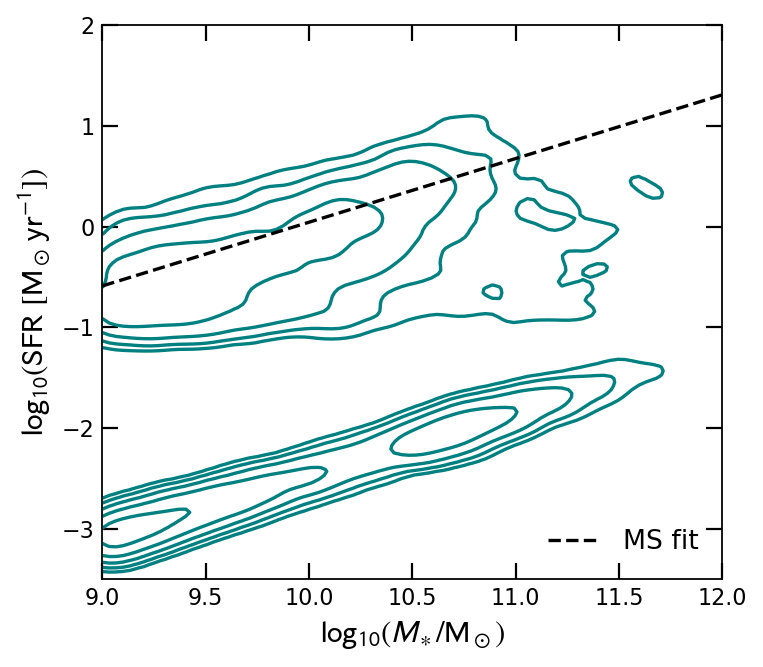}
\caption{$M_\ast-\rm{SFR}$ diagram of the galaxy sample in IllustrisTNG. The black
dashed line represents the Main Sequence fit (Eq.~\ref{eq:MS-TNG}),
while the logarithmically spaced contour lines indicate object 
density in the plane. In order to show the whole galaxy population,
all subhalos with SFR=0 are assigned a nominally low SFR 
value by taking a random draw for their $\Delta_{MS}$ from a~normal
distribution of $N(-2.5, 0.1)$.
\label{fig:msfr_TNG}}
\end{figure}

\subsection{Distance from the star-forming Main Sequence}
\label{sec:DeltaMS}
For each galaxy in the entire MaNGA sample, we calculate its distance from the star-forming Main Sequence (MS) as
\begin{equation}
\Delta_{MS} = \mathrm{\log_{10}(SFR \ [M_{\odot}/yr])} - \mathrm{MS}(M_{*}),
\label{eq:delta-ms}
\end{equation}
where 
\begin{equation}
\begin{split}
{\rm MS}(M_{*}) = -7.96 +  0.76\log_{10}(M_{*}[M_{\odot}])
\end{split}
\label{eq:msfr-manga}
\end{equation}
is our best fit of the star-forming MS, as presented in Fig.~\ref{fig:f_yp}. To perform this fit, we consider only galaxies with $\mathrm{\log_{10}(sSFR [1/yr])}>-11.5$, and perform a simple linear fit to all star-forming galaxies. For stellar masses $M_{*}$ and SFRs we use the values provided by the Pipe3D MaNGA value-added catalogue from \citet{Sanchez2016_pipe3d}, which are calculated from integration within the FoV of the datacubes, without further aperture correction. The integrated SFRs from Pipe3D are based on the well-established H$\upalpha$ calibration \citep{Kennicutt1998ARA&A..36..189K}, tracing star-formation on short timescales of $\sim 10$ Myr. The integrated H$\upalpha$ flux is dust corrected spaxel by spaxel  based on the Balmer decrement and the \cite{Cardelli1989ApJ...345..245C} dust attenuation law.

In IllustrisTNG, we also calculate $\Delta_{MS}$ with Eq.~\ref{eq:delta-ms} for all galaxies
in the sample. To do that, we follow the method adopted by \cite{Donnari2019}, performing 
a~linear fit to converged median SFR values of all galaxies with non-zero SFR, in bins of 0.2~dex in 
stellar mass, in the range \mbox{$9 < \log_{10}(M_\ast / \rm M_\odot) < 10.6$}. This yields
an $\rm MS_{\rm TNG}$ of the form:
\begin{equation}
{\rm MS}_{\rm TNG}(M_\ast) = -6.28 + 0.63\, \log_{10}(M_\ast\, [\rm M_\odot]),    
\label{eq:MS-TNG}
\end{equation}
shown as a black dashed line in Fig.~\ref{fig:msfr_TNG}.
We note that our definition of $\rm MS_{\rm TNG}$ has a shallower slope and a higher intercept 
than other studies \citep[e.g.][]{Donnari2019, Piotrowska2022}. This discrepancy
results from our choice of a different SFR averaging timescale and the applied conversion 
between Chabrier and Salpeter IMFs.

\subsection{Spectral stacking and fitting} \label{methodology}
For each MaNGA galaxy, we discard all the spaxels with a median $S/N<1$ per spectral pixel within the rest-frame wavelength range 4000–4500 Å. Then, we bin the spatially resolved MaNGA data in radial annuli (i.e. elliptical bins in the observed 2D projection of each galaxy, taking the viewing angle into account) using an adaptive scheme in order to ensure a high enough median S/N ($>35$) per spectral pixel for our purposes (the reason for this approach is to be consistent with our radial gradient analysis - we will discuss radial trends discovered with our methodology in a separate paper). 

The resulting binned spectra (summed without any weighting) are then fitted with our customised methodology, which is based on the $\chi^2$-minimization Penalized PiXel-Fitting code\footnote{\url{ https://pypi.org/project/ppxf/}; version 8.0.2} pPXF \citep{Cappellari2017, Cappellari2023}. To fit gas emission lines, we use Gaussians. To fit the stellar continuum, we use a library of simple stellar-population (SSP) templates coupled with a 10\textsuperscript{th}-order multiplicative Legendre polynomial. The SSP spectra are the synthetic spectra from the C3K library \citep{Conroy2019} using the MIST isochrones of \citet{Choi2016}, solar abundances, a Salpeter IMF, and a resolution of R=10,000. We use a total of 484 synthetic SSP spectra uniformly spanning the full 2D logarithmic grid of 44 ages and 11 metallicities from age$_{SSP}$ = $10^{6.0}$~yr to $10^{10.3}$~yr and [M/H] = -2.0 to 0.5, respectively. The large advantage of the C3K library is its expansion into young SSP ages in combination with its complete metallicity grid\footnote{These SSP templates are available from the author C. Conroy upon reasonable request.}. 

By fitting the observed spectra simultaneously with a superposition of SSPs and gas templates, we can reconstruct the complex formation and evolution history of the stellar system under consideration with this `astro-archaeological' approach, where we use the observed remaining stellar populations as the `fossil records' of the system's star-formation history (SFH). 

In more detail, our method, which we apply on each individual MaNGA galaxy in the selected sample (see Section \ref{sample_selection}), works as follows:
\begin{enumerate}
    \item First, the sky emission lines in each spectrum are masked; the templates are broadened to match the wavelength-dependant spectral resolution of the spectrum; and both the spectrum and the templates are re-normalized by the median flux per spectral pixel in the spectrum to avoid numerical issues, and to enable the use of regularization in pPXF, allowing to penalize non-smooth weight distributions (we use a second-order regularization, see \citealp{Cappellari2017,Cappellari2023} for more details). 
    \item Then, an initial pPXF fit is performed with a small regularization parameter of \citetext{regul = 5} in order to estimate the intrinsic noise of the spectrum and to exclude bad spectral pixels from the spectrum that might bias the fit. We estimate the noise $\sigma(\lambda)$ in the spectrum as function of wavelength $\lambda$ via a running average of the 16\textsuperscript{th}\&84\textsuperscript{th}-percentile residuals within $\pm 50$ spectral pixels for each pixel in the spectrum.
    \item Next, we perform a 3-$\sigma$ clipping on the spectrum. 
    As the flux of gas emission lines often exceeds the underlying flux of the stellar continuum by multiple factors or even order of magnitudes, and with that also their residuals are often substantially larger, gas emission lines are excluded from $\sigma$ clipping. Typically, 1--3~per cent of spectral pixels are excluded by $\sigma$ clipping.
    \item We fit the spectrum again with pPXF, excluding the $\sigma$-clipped pixels and using our updated noise estimate, again with regul = 5. This initial pPXF fit with best-fitting solution $y(\lambda)$ is the basis function of our running bootstrapping method described in the next items.
    \item We perform a residual-based bootstrapping by perturbing $y(\lambda)$ with the the residuals $\epsilon(\lambda) = y(\lambda) - s(\lambda)$ of the initial fit: $y^*(\lambda) = y(\lambda) \pm \epsilon(\lambda')$, where $s(\lambda)$ is the observed galaxy spectrum and $\epsilon(\lambda')$ is randomly chosen from all residuals within $\pm 50$ pixels for each spectral pixel. 
    \item We fit the perturbed spectrum $y^*(\lambda)$ again with pPXF, this time without regularization (regul = 0).
    \item We iteratively repeat steps 4\&5 one hundred times.
    \item We average each grid point of the age--metallicity weight grid over all iterations to recover a SFH consistent with the intrinsic noise of the spectrum. This method probes the sampling distribution of each individual SSP grid weight. 
\end{enumerate}

 One of the key advantages of our approach is that the fitted weights of the 484 SSP templates are independent from one another (apart from the little regularization we use to slightly smooth the weight distribution, see \citealp{Cappellari2017} for more details). Hence, our recovered SFHs are non-parametric and do not depend on any assumption about the underlying physics of galaxy evolution. Crucially, any recovered scaling relation cannot have been introduced by parametric assumptions about the shape of our fitted 2D SFHs.

While this non-parametric astro-archaeological approach is extremely powerful, the analysis of its outputs has to be done with utmost care. Depending on the $S/N$ of a considered spectrum, the 2D-shape of the recovered SFHs (e.g. due to the well known age-metallicity degeneracy), and known and unknown systematics, such as dust obscuration,  or flux calibration issues, we can trust returned SSP weights to different degrees. To assess the stability of the SSP weights, the bootstrapping methodology described above is highly instructive, as it returns a scatter distribution for each individual SSP weight, as well as for any quantity derived from the weight grid; e.g. the light-weighted metallicity, given as the weighted average  metallicity of all fitted SSPs. The thorough tests performed to assess the validity and stability of our results discussed in this paper will be discussed in detail in Looser et~al. (in prep.). Our tests showed that numerous interesting, `collapsed' (i.e. averaged over many galaxies and radial bins therein) quantities derived from a large sub-sample of SSPs are highly reliable. 

To give an example of a collapsed quantity derived from the full spectral fitting, Fig.~\ref{fig:f_yp} shows the fraction of total fitted light stemming from all SSPs younger than $10^{8.5}$\ yr, $f_{l,YP}$, averaged over all galaxies with similar stellar masses and SFRs. Contributing to this plot are all galaxies in the MaNGA sub-sample used in this work, see Section \ref{sample_selection}. The reader will note the small scatter in $f_{l,YP}$ between neighboring hexagonal bins, as well as the general agreement with physical expectations between $f_{l,YP}$ and the location of galaxies on the SFR--$M_*$ plane.

\subsection{Sample selection} \label{sample_selection}
For the purposes of this analysis, we only consider MaNGA galaxies which have an integrated median $S/N>35$ per spectral pixel in the wavelength range 4000--4500 Å (in rest-frame). This ensures the reliable inference of the quantities presented in this paper \citep[see e.g.,][]{barone+2020}. Further, we discard targets classified as AGN in the BPT diagram and discard 20 galaxies for which our fitting methodology failed due to flux calibration or contamination issues. This results in the sub-sample of 7323 MaNGA galaxies used in this work.

As can be seen in Fig.~\ref{fig:f_yp}, the selected sample densely populates both the MS and the quiescent population (each hexagonal bin comprises at least 15 galaxies) over a wide range of stellar masses.

In IllustrisTNG we select all subhalos of cosmological origin (i.e. genuine structures 
with \texttt{SubhaloFlag}=1, as opposed to spurious Subfind algorithm identifications)
with $M_\ast > 10^9 \,  \rm M_\odot$, to match simulated and observed galaxy samples
in mass. Additionally, we remove all satellite galaxies with dark matter mass fraction 
lower than 10~per cent, to confidently eliminate subhalos potentially misidentified with the
SubFind algorithm \citep{Pillepich+2018b,Genel+2018}. These combined selection criteria yield 
a total sample size of 17~206 individual galaxies.


\section{The stellar Fundamental Metallicity Relation} \label{sec:FMR}

\subsection{The light-weighted stellar Fundamental Metallicity Relation in local galaxies from MaNGA}

\begin{figure}
\includegraphics[width=1.0\columnwidth]{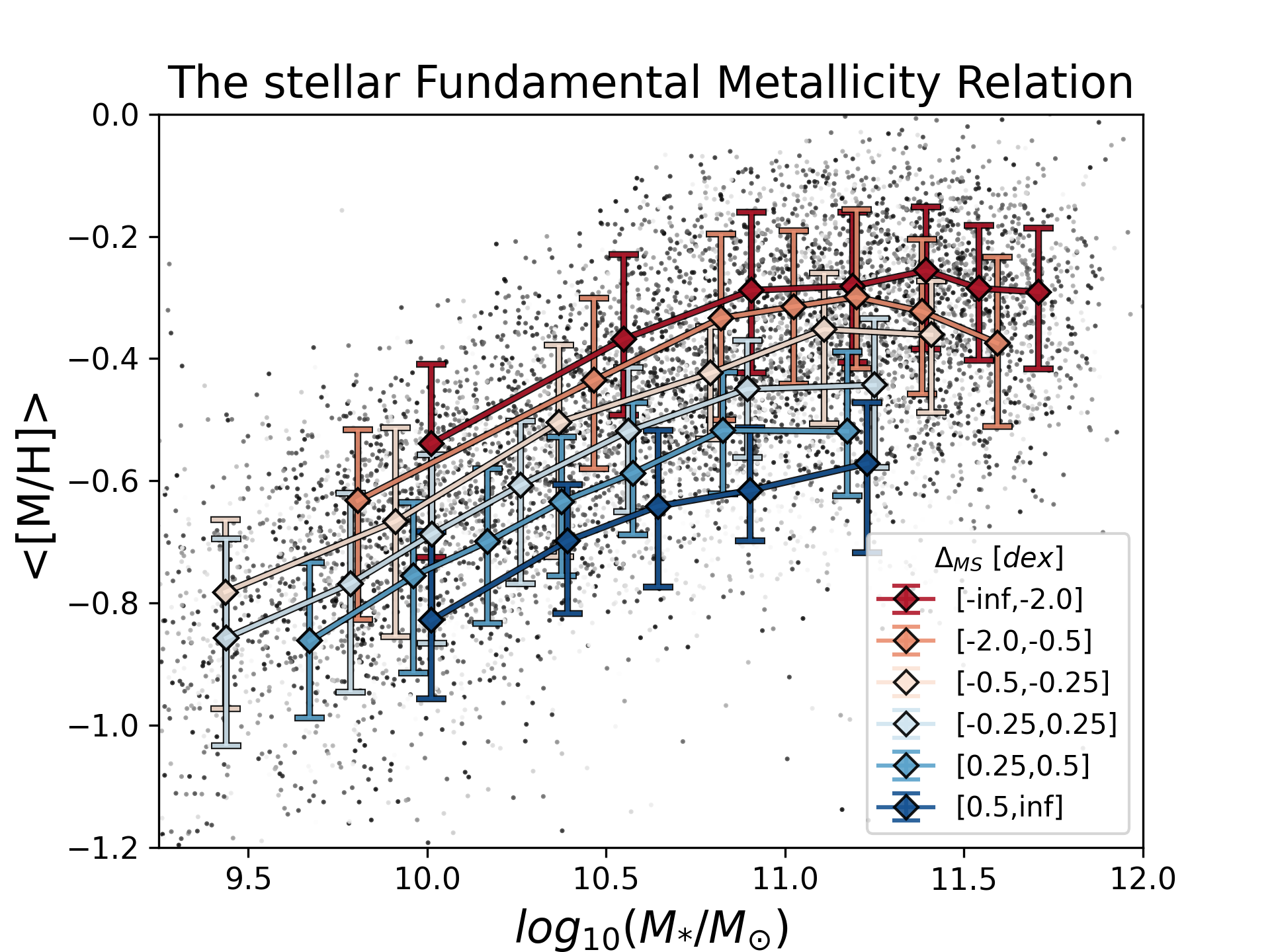}
\caption{The global light-weighted stellar Fundamental Metallicity Relation (sFMR) in local galaxies from MaNGA. Each grey dot represents one galaxy.
The diamonds represent the median integrated stellar metallicity in each $\Delta_{MS}$-$M_{*}$-bin, while the error-bars indicate the 16th and 84th percentiles. Each $\Delta_{MS}$-$M_{*}$-bin comprises at least 80 galaxies. 
\label{fig:MaNGA_FMR}}
\end{figure}

In  Fig.~\ref{fig:MaNGA_FMR} we present the global stellar Fundamental Metallicity Relation (sFMR). For this purpose, we collapse our spatially resolved results into a global analysis, simply by summing the results over the annular bins in each galaxy. The stellar metallicity of each galaxy \mhavg is defined as the light-weighted average metallicity of all SSPs fitted with the methodology described in Section \ref{methodology}. Further, we split the selected sample of 7323 MaNGA galaxies into six $\Delta_{MS}$-bins, as indicated in the legend of Fig.~\ref{fig:MaNGA_FMR}. In addition to the $\Delta_{MS}$-binning, we apply an adaptive $M_{*}$ binning in each $\Delta_{MS}$ bin to create a $\Delta_{MS}$--$M_{*}$ 2D binning of the sample with at least 80 galaxies in each 2D bin.

It is evident that different local $\Delta_{MS}$-galaxy populations follow different mass--metallicity relations (MZRs), where the higher $\Delta_{MS}$ populations are systematically metal poorer than the more quiescent populations; at all stellar masses and between all $\Delta_{MS}$ bins. While the MZR is steeply increasing for less massive systems in all $\Delta_{MS}$ bins, a flattening is evident at the high-mass end for galaxies on and below the MS. 
In summary, we clearly identify a sFMR in our non-parametric SSP analysis of the MaNGA data, analogous to the well known gFMR.

\subsection{The mass-weighted stellar Fundamental Metallicity Relation in IllustrisTNG}
\label{sec:sFMR_IllustrisTNG}

\begin{figure}
    \centering
    \includegraphics[width=\columnwidth]{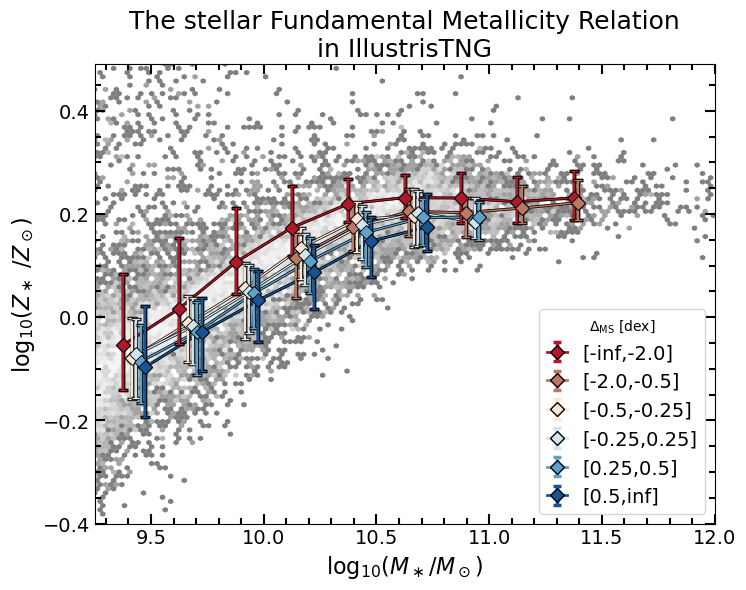}
    \caption{Global mass-weighted stellar FMR in IllustrisTNG. 
    The diamonds show median mass-weighted stellar metallicity in each $\Delta_{MS}$--$M_\ast$ bin,
    while the error-bars correspond to the 16\textsuperscript{th} \& 84\textsuperscript{th} percentiles of the
    associated $Z_\ast$ distribution. Each of these bins comprises at least 40 galaxies
    and $\Delta_{MS}$ bins are horizontally offset to improve figure clarity.
    Hexagonal bins indicate the density of objects in the plane.}
    \label{fig:mzr-TNG}
\end{figure}

In Fig.~\ref{fig:mzr-TNG} we present the stellar FMR extracted from the IllustrisTNG 
cosmological simulation. Similarly to our MaNGA analysis, we split the simulated sample 
in ranges of $\Delta_{MS}$ identical to Sec.~\ref{sec:FMR}, and within each 
$\Delta_{MS}$ range, we further divide objects into bins of 0.25~dex 
in $M_\ast$. Each colored diamond indicates a median value
in a~bin with at least 40 galaxies, while error-bars indicate the 16\textsuperscript{th} \& 84\textsuperscript{th} percentiles of the $Z_\ast$ distribution in said bin. Greyscale
hexagonal bins indicate the density of objects in the plane.

Fig.~\ref{fig:mzr-TNG} shows a well-defined relationship between 
total stellar mass of simulated galaxies and the average metallicity of 
their constituent stars. The simulated sFMR evolves
modestly by $\sim 0.3\ \rm dex$ over 2.5~dex in $M_\ast$, an evolution 
similar to that of the gFMR reported in \cite{Torrey2019}, albeit with a smaller 
apparent scatter. Comparing trends among individual mass bins, we find that all 
$\Delta_{MS}$ bins follow similar trajectories in the $M_\ast - Z_\ast$
plane, which are vertically offset in a direction of higher metallicity 
associated with highest negative deviation from the star-forming Main Sequence.
This trend agrees well with our analysis in Sec.~\ref{sec:FMR}, however
appears less pronounced owing to the tight nature of the
simulated sFMR. 

At this point we stress that \textit{one cannot compare the absolute values 
of light-weighted stellar metallicities} estimated in the observations
\textit{with mass-weighted metal fractions} calculated for stellar particles
within cosmological hydrodynamical simulations. As demonstrated
in \cite{Nelson2018}, in the absence of full forward-modelling 
and observation-like spectral fitting, stellar metallicities estimated 
in simulations are offset to higher values with differences reaching 
up to $\sim 0.5\ \rm dex$. Even if complex forward modelling is applied to simulate observations, the light-weighted metallicity values
recovered from simulated galaxies are likely to show quantitative deviations from
their observed counterparts (see Appendix~\ref{Appendix:lw_sFMR_Illustris}).
On the other hand, mass-weighted metallicities for local galaxies are unreliable due to outshining and to degeneracies between the oldest stellar populations (Appendix~\ref{Appendix:mw_sFMR}).
For this reason, our main focus lies in a 
qualitative comparison between relative trends in $Z_\ast$, rather than its absolute values.

To attempt a more direct comparison, the observed mass-weighted sFMR and the simulated light-weighted sFMR are presented in Figs.~\ref{fig:MaNGA_FMR_mw}
and~\ref{fig:tng-stellar-fmr-lightweighted}. 
A direct, quantitative comparison between observations and simulations is beyond the scope of this article.

\subsection{The Fundamental Metallicity Relation in young stellar populations}
\begin{figure}
\includegraphics[width=1.0\columnwidth]{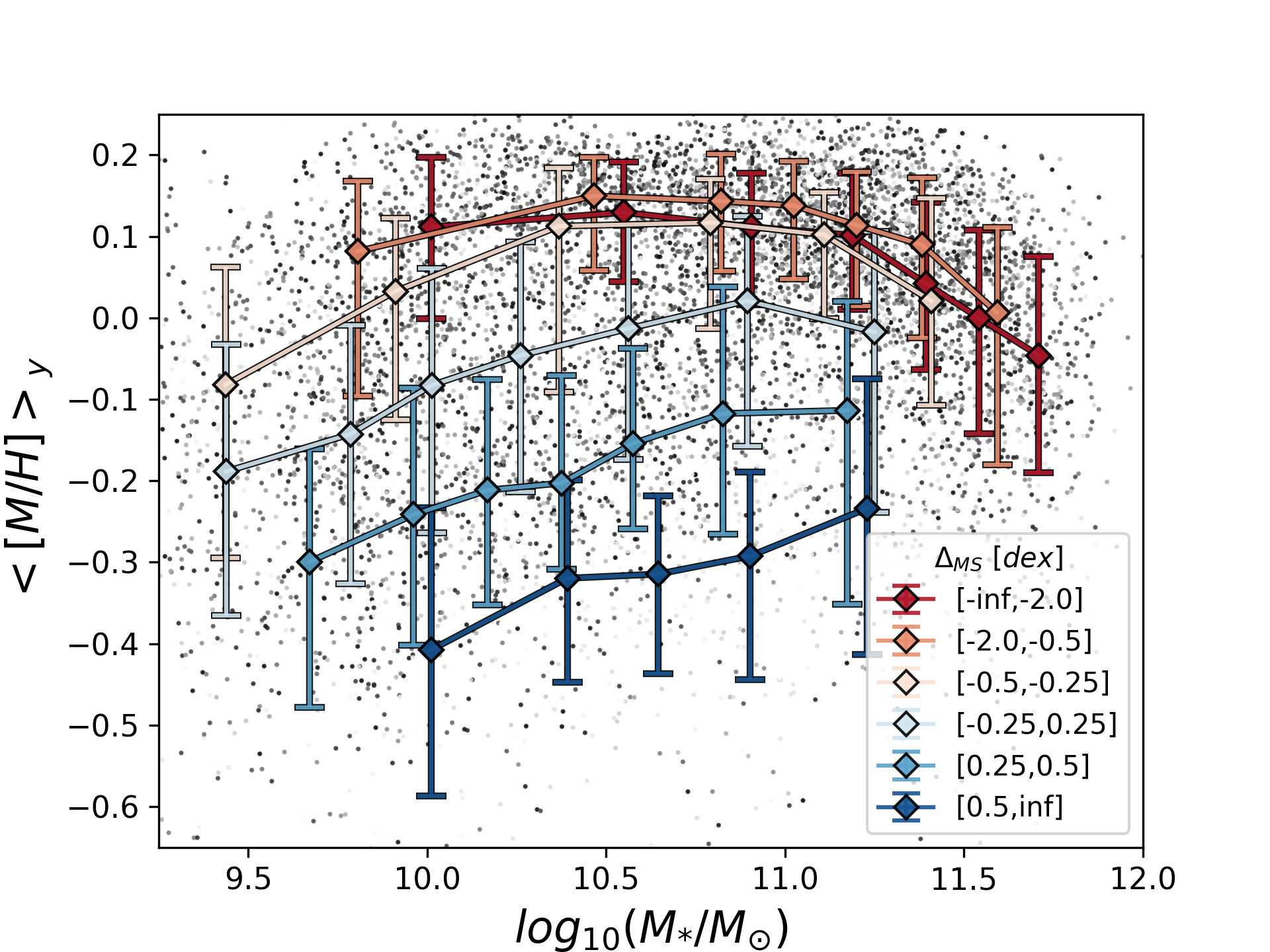}
\caption{The stellar Fundamental Metallicity Relation (sFMR) for young stellar populations in MaNGA. Each dot represents one galaxy. 
The diamonds represent the median integrated stellar metallicity in each $\Delta_{MS}$-$M_{*}$-bin, while the error-bars indicate the 16 and 84 percentiles dispersion. 
\label{fig:MaNGA_FMR_y}}
\end{figure}

In  Fig.~\ref{fig:MaNGA_FMR_y} we present the FMR of the young stellar populations in local galaxies, as recovered by our non-parametric methodology described in Section \ref{methodology}. The ``young metallicity' of each galaxy is defined as the light-weighted average metallicity of all SSPs younger than $10^{8.5}$\ yr. We use the same $\Delta_{MS}$--$M_{*}$ 2D binning as in section \ref{sec:FMR}, with at least 80 galaxies in each bin. Analogous to the sFMR, galaxies in different $\Delta_{MS}$ bins follow different young SSP mass-metallicity relations: at all stellar masses and between all $\Delta_{MS}$-bins, the more star-forming populations are systematically metal poorer than the more quiescent populations. The young metallicity values are systematically higher than the total metallicity values for galaxies at all masses and distances from the MS, showing that we recover the expected chemical enrichment of galaxies with cosmic time, at least qualitatively. This serves as an independent validation of the methodology, because our non-parametric approach allows \emph{any} chemical-evolution history. A detailed study of the chemical enrichment of individual galaxies is presently beyond the scope of this work.

Quiescent galaxies show a negative mass--metallicity slope at high stellar mass. Due to the very small amount of flux stemming from young SSPs in these massive, quiescent systems (as one arguably expects), see Fig.~\ref{fig:f_yp}, this particular trend likely cannot be trusted. 
A detailed analysis of the reliability of the recovery of metallicity for young stellar populations will be presented in future work.

However, if confirmed, the young sFMR consistently links the timescales between the well established gFMR and the stellar FMR in local galaxies presented in this work.

\section{Discussion} \label{sec:Discussion}
It is now a well established result that younger galaxies (as measured by their average stellar ages) have systematically lower metallicity compared to older galaxies \citep{Scott2017MNRAS.472.2833S, Li2018MNRAS.476.1765L, Neumann2021MNRAS.508.4844N}. And recently, it has been shown that stellar metallicities of local galaxies show a \emph{continuous} trend with age \citep{Cappellari2023}, tracing the complex star-formation history of galaxies over cosmic times. 

In this work, instead, we connected the chemical-enrichment histories of galaxies to their star-formation activity in the recent past, by presenting evidence for the existence of a stellar FMR, analogous to the well-known gas-phase FMR. This is evidence for the chemical evolution of the ISM being imprinted onto the chemical evolution history of the stars over cosmic times. In addition to connecting the gas and stellar metallicities, we also `resolve' the gap in stellar metallicity between star-forming and quiescent galaxies. More concretely, we show that there is a \emph{continuous} distribution of metallicities, which correlates with the distance from the MS in both observed local galaxies in MaNGA and equivalent simulated galaxy populations in IllustrisTNG. 


Additionally, we find strong evidence that a stellar FMR also exists in the young ($<$300~Myr) stellar populations of local galaxies, despite these populations being systematically more metal rich than the older populations. Thus, we find that our non-parametric star-formation and chemical-enrichment histories require (i) the metallicity of individual galaxies to increase with cosmic time and (ii) the total metallicity to be progressively larger, moving from above the MS towards the passive galaxies.

\subsection{The starvation hypothesis}
\citet{Peng2015} and \citet{Trussler2020} interpreted the `metallicity gap' between local star-forming and quiescent galaxies as evidence for so-called `starvation', i.e. the continuation of star formation after the supply of low-enrichment gas has been interrupted. Starvation causes efficient enrichment of the ISM because -- by definition -- it requires repeated recycling of the star-forming gas without diluting it with low-enrichment, newly-accreted gas. The underlying assumption of the starvation hypothesis is that -- eventually -- low-metallicity/high-$\Delta_{MS}$ galaxies will evolve in high-metallicity/low-$\Delta_{MS}$ galaxies.
Admittedly, we are comparing the metallicities of different $\Delta_{MS}$ bins in the local Universe, corresponding to different galaxies which are not linked by progenitor-descendant relations in Figs.~\ref{fig:MaNGA_FMR} -- \ref{fig:MaNGA_FMR_y}. However, if we were to compare the stellar metallicities of quiescent galaxies to the metallicities of their star-forming progenitors, the gap would be even wider \citep{Peng2015, Trussler2020}.

\subsection{The metal retention hypothesis}
In contrast to the starvation hypothesis, \citet{Vaughan2022} argue that the offset between the MZRs of star-forming and passive galaxies is mostly due to their differing gravitational potential wells, for which they use 
$M_{*}/R_{\rm e}$ 
as a proxy: everything else being equal, galaxies with
higher $M_{*}/R_{\rm e}$ might be able to retain more metals by reducing metal-loaded outflows.
Indeed, analogously to \citet{Vaughan2022}, we find that the 
scatter in the \mhavg--$M_{*}/R_{\rm e}$ relation 
is reduced relative to the \mhavg--\mstar relation (Fig.~\ref{fig:M_re_FR}, where $R_{\rm e}$ is taken from the Pipe3D MaNGA value-added catalogue \citep{Sanchez2016_pipe3d}).
We note that $M_{*}/R_{\rm e} \equiv 5 \sigma_{\rm vir}^2 / G \approx 5 \sigma_{\rm e}^2 / G$ is equivalent to the virial estimate $\sigma_{\rm vir}$ of the stellar velocity dispersion $\sigma_{\rm e}$; this means that the reduced scatter going from $M_{*}$ to $M_{*}/R_\mathrm{e}$
is also consistent with previous works, which showed that $\sigma_\mathrm{e}$ is better than $M_{*}$ as a predictor of stellar population properties
\citep{kauffmann+2003a, franx+2008, Cappellari2011dur, cappellari+2013b} like colour
\citep[e.g.,][]{bell+2012, wake+2012}, metallicity \citep{barone+2018, barone+2020, barone+2021} or both age and metallicity \citep{McDermid2015MNRAS.448.3484M,Cappellari2016ARA&A..54..597C,Li2018MNRAS.476.1765L, Cappellari2023,Lu2023}.

In particular, the three lowest-$\Delta_{MS}$ bins follow a nearly identical \mhavg--$M_{*}/R_{\rm e}$ relation. However, the scatter we find between individual $\Delta_{MS}$ bins on or above the MS is still significant and the offset between the most star-forming to the most quiescent bin only reduces from ca. 0.3-0.4 dex to ca. 0.2-0.3 dex, when exchanging \mstar with $M_{*}/R_{\rm e}$ on the x-axis. 
The difference between our results and \citet{Vaughan2022} is most likely due to the different samples used. Compared to SAMI, MaNGA contains more high-\mstar star-forming galaxies, enabling us to measure the metallicity of massive galaxies on and above the MS.

When considering $M_{*}/R_{\rm e}$ instead of \mstar, the metallicity offset between star-forming and quiescent galaxies reduces by $\approx$0.1~dex.
Assuming that $M_*/R_\mathrm{e}$ is a good proxy for the gravitational potential \citep[as proposed by e.g.,][]{Vaughan2022}, the observed reduction in scatter going from the mass--metallicity to the potential--metallicity relation could be explained by the fact that $M_*/R_\mathrm{e}$ encapsulates the capability of galaxies to retain metals (by resisting outflows). However, the decrease in scatter by 0.1~dex is only 25~per cent of the spread in metallicity between different $\Delta_{MS}$ bins.
This fact indicates that -- even though the capability of galaxies to retain metals by preventing outflows is having an effect, other physical processes are likely dominant in forming the stellar FMR. 

\subsection{The Fundamental Metallicity Relations: correlating star formation over long timescales}
The gFMR connects star-formation and gas metallicity on timescales that are at least as long as the timescale probed by the SFR indicators (e.g. $\approx$10~Myr for H$\upalpha$). This is consistent with short-lived `episodic' accretion of gas,
capable of lowering the metallicity of the ISM and increasing the SFR \citep[e.g.][]{Almeda2014A&ARv..22...71S, Almeida2017ASSL..430...67S}.
The sFMR enables us to extend this correlation to a timescale of order $1/sSFR$, because galaxies have to form enough new stars to change the total light-weighted metallicity, including the extant stellar populations. This means that the FMRs are not due to short-lived `episodic' accretion, but to long-lasting inflow of low-metallicity gas from the IGM/CGM, simultaneously fueling star formation and
keeping the ISM in a chemical-enrichment equilibrium between metal production and dilution. This finding is in agreement with theoretical models \citetext{see e.g. \citealp{forbes+2014} for IGM accretion and \citealp{Torrey2019} for CGM accretion}. Our interpretation can also explain the gFMR, is consistent with the existence of a `gas FMR' \citep[i.e., the relation between metallicity, stellar mass and gas mass,][]{Bothwell2013}, and agrees with the observation that (central) galaxies lying below the MS have lower gas fractions \citetext{\citealp{Baker2022b, Tacconi2020, Saintonge2016} -- with star-formation efficiency playing at most an equal role \citealp{Piotrowska2020}}. We also note that the existence of a long-timescale correlation for star-forming galaxies does not contradict the presence of short-duration bursts \citep[e.g.][]{Wang2019}, because the two can co-exist \citep{Caplar2019, Tacchella2020MNRAS.497..698T}.

\subsection{Starvation drives galaxy quenching}
\begin{figure}
\includegraphics[width=1.0\columnwidth]{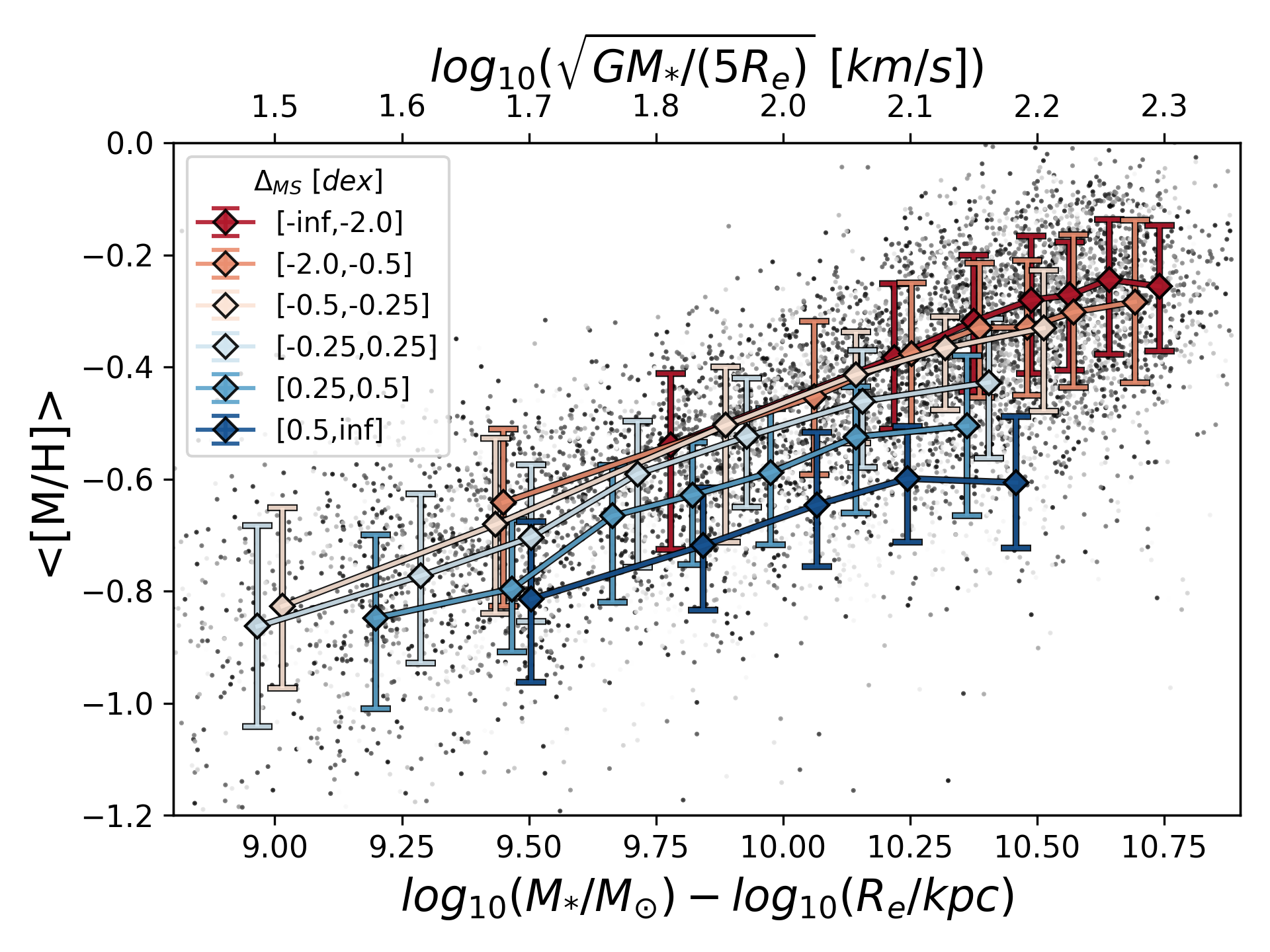}
\caption{The difference in average stellar metallicities of different $\Delta_{MS}$-populations as a function of 
$M_{*}/R_{\rm e}$, equivalent to the the viral estimate of the velocity dispersion: $\sigma_{\rm vir}^2 \equiv G M_{*}/5 R_{\rm e}$. 
Each dot represents one galaxy.
The diamonds represent the median integrated stellar metallicity in each $\Delta_{MS}$-$M_{*}/R_{\rm e}$ bin. The error-bars indicate the 16 and 84 percentiles dispersion.}
\label{fig:M_re_FR}
\end{figure}

The anti-correlation between stellar metallicity and distance from the MS at all masses shows that galaxies increase their stellar metallicity as they leave the MS. If this was not true, we should be observing low-metallicity galaxies on their way to quiescence and in the quiescent population, but this is not supported by observational evidence: the 1-$\sigma$ scatter of the quiescent stellar MZR is only $\approx$0.1~dex, inconsistent with a large population of local low-metallicity quiescent galaxies lying 0.2--0.3~dex below it. The absence of such galaxies and the independent observation of lower gas fractions below the MS \citep{Saintonge2016, Baker2022, Baker2022b} suggest that galaxy quenching is accompanied by starvation. 

The three $\Delta_{MS}$ bins below the MS show very similar \mhavg--$M_{*}/R_{\rm e}$ relations, with only a very small residual dependence on $\Delta_{MS}$ (red hues in Fig~\ref{fig:M_re_FR}). In contrast, the \mhavg--$M_{*}/R_{\rm e}$ relations of the three bins on or above the MS still exhibit a strong dependence on $\Delta_{MS}$ (blue hues in Fig.~\ref{fig:M_re_FR}). This difference suggests that different physical mechanisms set the stellar metallicities in star-forming and quiescent galaxies.


The galaxies below the MS are either quiescent or on their way to quiescence; this
means that they are not going to form a significant amount of stars relative to their extant stellar mass, which in turn means that it is unlikely they are going to significantly change
their current stellar metallicity, which strongly correlates with $M_*/R_{\rm e}$. On the other hand, at all $M_{*}/R_{\rm e}$, the metallicity of star-forming galaxies anti-correlates with $\Delta_{MS}$. Recalling that $\Delta_{MS}$ is primarily related to the gas fraction, this observation suggests that for star-forming galaxies
stellar metallicity is set jointly by $M_{*}/R_{\rm e}$ (or by $M_{*}$) and by the specific inflow rate of metal-poor gas. These systems are expected to significantly increase their stellar mass, therefore their stellar metallicity will evolve during the starvation process.

Coupled with independent strong evidence 
that quenching is \emph{causally connected} to the integrated supermassive black-hole feedback \citep{Piotrowska2022, Bluck2023}, the emerging scenario is one where AGN activity reduces and eventually interrupts the inflow of cold gas to the galaxy. During the extended phase of halted or significantly decreased accretion of pristine gas from the CGM/IGM, galaxies continue to consume gas, form stars and recycle gas via supernova explosions and stellar winds, which enriches their ISM - out of which stars with higher metallicities are born. During this long-lasting starvation process, the galaxies continuously decrease their gas reservoirs and transition through the different $\Delta_{MS}$ bins, while increasing their average metallicities -- driven by the increase in metallicities of their youngest stellar populations -- until they run out of gas, cease forming stars and end up on the red sequence. 

This process establishes the observed stellar FMR and the \mhavg--$M_{*}/R_{\rm e}$--$\Delta_{MS}$ relation in local galaxies. These relations therefore pose constraints to the duration of quenching itself. 



\section{Summary} \label{sec:Summary}

In this work we use integral-field spectroscopy from the local MaNGA survey coupled with a non-parametric SFH recovery methodology based on the full spectral fitting code pPXF, to connect the stellar mass, star-formation rate and stellar metallicity of a representative sample of local galaxies ($z<0.15$).

\begin{itemize}
    \item The main result of this work is the discovery of the stellar FMR, analogous to the well-established gas FMR. 
    \item When we compare the observed sample of MaNGA galaxies
    to an equivalent population in IllustrisTNG, we find matching relative
    trends in $Z_\ast$ and $M_\ast$. The absolute values are offset, however,
    due to the intrinsic differences between methods employed in measuring
    metal content in simulations and observations.
    \item Similarly to the overall stellar FMR, we find a young stellar FMR, when considering the average metallicity of only the SSPs younger than $10^{8.5}$\ yr in local galaxies. 
    \item As stars continuously form from the gas in the ISM, we propose that the young stellar FMR and the total stellar FMR are both a natural consequence of the gFMR (which is valid to at least z=3), which continuously gets imprinted on the stellar populations over cosmic epochs. Hence, the stellar, the young stellar, and the gFMR all trace the same physical processes, imprinted into the baryonic properties of galaxies over cosmic timescales, intermediate timescales, and timescales of ca. 10 Myr, respectively.

    \item The observed presence of a stellar FMR suggests that starvation is the dominant mode through which galaxies quench.
    
    \item The ability of galaxies to retain metals, by hindering outflows during their starvation phase, which might be correlated with $M_{*}/R_{\rm e}$,
    can partially explain the differences in MZRs between different $\Delta_{MS}$ bins. However, even after incorporating this information, the metallicity differences are still significant.

    \item The continuous accretion of metal-poor (or even pristine) gas onto star-forming galaxies, still ongoing in the local Universe, plays an important role in forming all three, the stellar FMR, the young FMR and the gFMR. The continuous accretion of metal-poor gas onto star-forming galaxies - over cosmic epochs - simultaneously provides the fuel for ongoing star formation and continuously dilutes the ISM, out of which new stars are formed. 
    
    
    
\end{itemize}

\

The existence of the stellar FMR leads to the conclusion that the total star formation of galaxies is correlated on long timescales (at least in the low-redshift Universe), and that the stellar metallicity of local quiescent galaxies is largely set during quenching via starvation, i.e. a slow transition to quiescence due to the suppression of cold-gas accretion.


\section*{Acknowledgements}

T.J.L., F.D’E., J.M.P., R.M. and W.B. acknowledge support by the Science and Technology Facilities Council (STFC), by the ERC through Advanced Grant 695671 ``QUENCH", and by the
UKRI Frontier Research grant RISEandFALL. T.J.L acknowledges support by the STFC Center for Doctoral Training in Data Intensive Science. R.M. also
acknowledges funding from a research professorship from the Royal Society. 
We thank Simon Lilly and Asa Bluck for very useful discussions.

\section*{Data availability}
This paper uses data from the Mapping Nearby Galaxies at Apache Point Observatory (MaNGA) survey \citep{Bundy2015}, which is part of the fourth-generation Sloan Digital Sky Survey (SDSS-IV). We use MaNGA DR17, which is s publicly available on https://www.sdss4.org/dr17/manga/. All extracted quantities used in this work will be made available as supplementary data upon acceptance of the paper.

%




\appendix

\section{The mass-weighted stellar FMR} \label{Appendix:mw_sFMR}

\begin{figure}
\includegraphics[width=1.0\columnwidth]{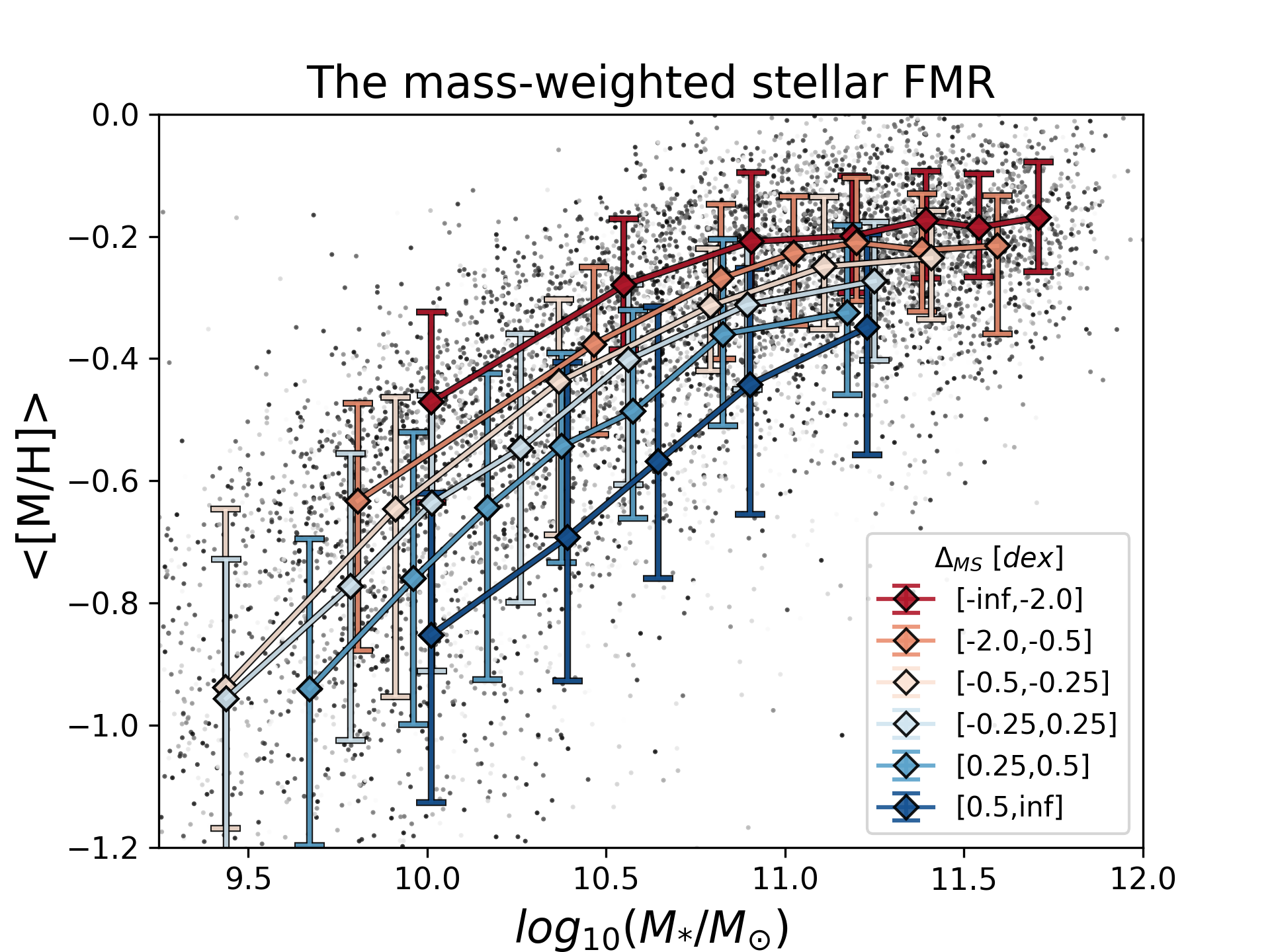}
\caption{The global mass-weighted stellar Fundamental Metallicity Relation (sFMR) in local galaxies from MaNGA. Each grey dot represents one galaxy.
The diamonds represent the median integrated stellar metallicity in each $\Delta_{MS}$-$M_{*}$-bin, while the error-bars indicate the 16th and 84th percentiles. Each $\Delta_{MS}$-$M_{*}$-bin comprises at least 80 galaxies. 
\label{fig:MaNGA_FMR_mw}}
\end{figure}

In  Fig.~\ref{fig:MaNGA_FMR_mw} we present the mass-weighted global stellar Fundamental Metallicity Relation. The mass-weighted stellar metallicity of each galaxy \mhavg is defined as the mass-weighted average metallicity of all SSPs fitted with the methodology described in Section \ref{methodology}. In other words, Fig.~\ref{fig:MaNGA_FMR_mw} is identical to Fig.~\ref{fig:MaNGA_FMR}, apart from mass-weighting the SSP grid before calculating the stellar metallicity of each galaxy. The observed mass-weighted FMR is qualitatively similar to the mass-weighed sFMR predicted by IllustrisTNG, presented in Fig.~\ref{fig:mzr-TNG}. Nonetheless, we caution against the over-interpretation of the observed mass-weighted stellar FMR, and particularly against mass-weighted metallicities inferred for individual galaxies. The old (and metal-rich) SSPs are hard to constrain 
due to the faintness of these stellar populations. This makes them susceptible to outshining by the younger, brighter populations. Or conversely, the fit may infer spurious old stellar populations, biasing the metallicity measurement. In light-weighted quantities this is not an issue, as the oldest templates only contribute little to the fit and only marginally to the inferred light-weighted SSP grid (typically less than 5 \%). However, these untrustworthy populations may contribute considerably to the mass-weighted SSP grid, making mass-weighted quantities less reliable. The introduction of spurious old, metal-rich populations in the pPXF-fits explains the dominantly increased normalisation of the mass-weighted sFMR in Fig.~\ref{fig:MaNGA_FMR_mw} compared to Fig.~\ref{fig:MaNGA_FMR}.

\section{The light-weighted stellar FMR in TNG50} \label{Appendix:lw_sFMR_Illustris}

\begin{figure}
\includegraphics[width=.9\columnwidth]{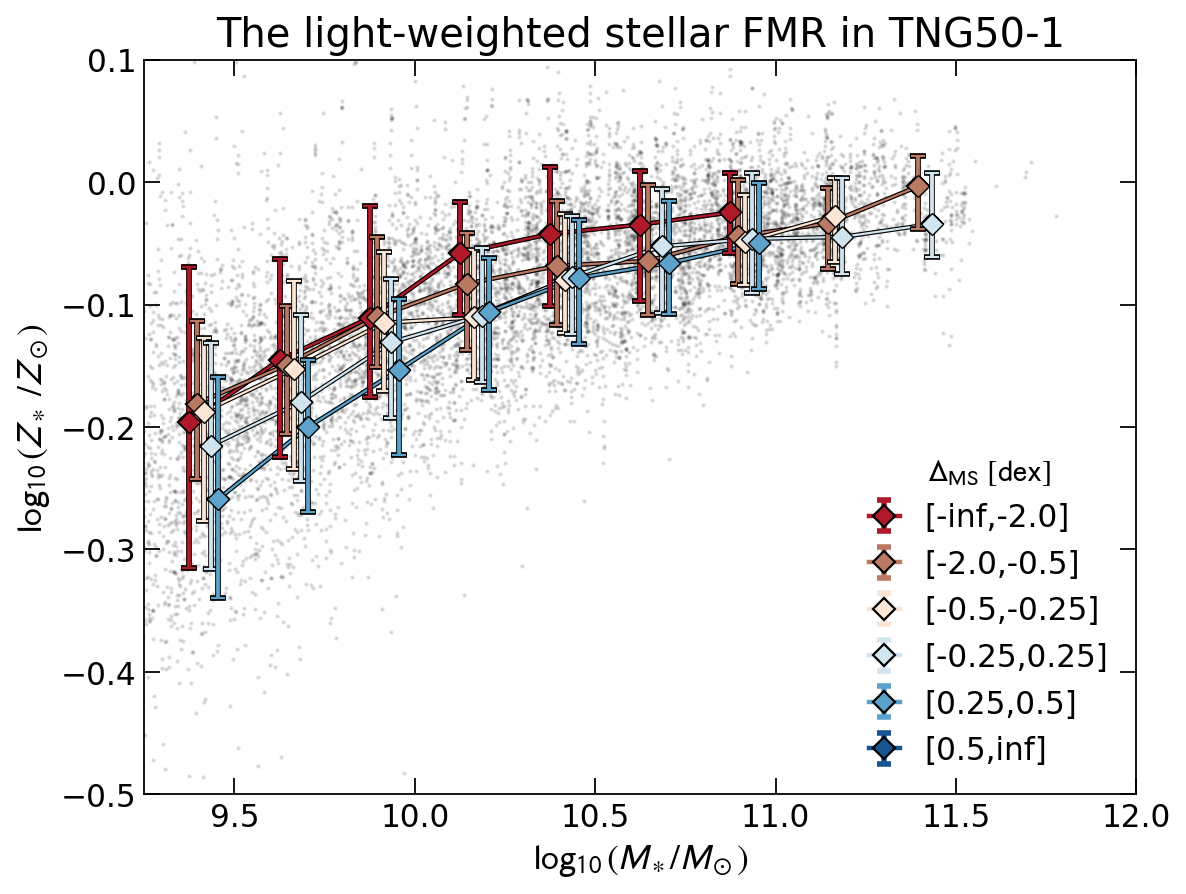}
\caption{The global light-weighted stellar Fundamental Metallicity Relation
in a sample of TNG50-1 galaxies selected to match the MaNGA sample
selection criteria \citep{Sarmiento2023}. Each grey dot represents one galaxy.
The diamonds represent the median integrated stellar metallicity in each \mbox{$\Delta_{MS}$-$M_{*}$-bin}, while the error-bars indicate the 16th and 84th percentiles. Each $\Delta_{MS}$-$M_{*}$-bin comprises at least 80 galaxies. 
\label{fig:tng-stellar-fmr-lightweighted}}
\end{figure}
To further investigate discrepancies between the stellar FMR
seen in observations and cosmological simulations, we use publicly available catalogues of MaNGA-like mock-galaxy observations 
generated from the highest resolution run of the Illustris-TNG suite, TNG50-1, performed for a volume of $(50\ \rm cMpc)^3$. In particular, we make use
of \textit{MaNGIA} 2D stellar property maps \citep{Sarmiento2023}
to extract global light-weighted mean metallicities for a sample of 
10,000 galaxies selected from the simulation to match the MaNGA survey
in terms of mass, size and redshift. Each mock galaxy observation 
was produced by assuming synthetic stellar templates based on the 
MaStar library \cite{Yan2019}, a uniform metallicity-dependent dust 
screen model and \textsc{CLOUDY} templates for \textsc{HII} region
emission associated with star-forming gas cells. Each mock was then 
analysed with pyPipe3D in a fashion identical to \cite{Sanchez2022}
to generate 2D maps of stellar properties including age, metallicity
and kinematics.

In Fig.~\ref{fig:tng-stellar-fmr-lightweighted} we present the 
light-weighted global stellar Fundamental Metallicity Relation
in the MaNGIA mock galaxy catalogue in TNG50-1. 
Fig.~\ref{fig:tng-stellar-fmr-lightweighted} is identical in spirit
to Fig.~\ref{fig:mzr-TNG}, however both the sample of galaxies and
the metallicity estimates are replaced. In order to calculate
$\Delta_{MS}$ values for the sample, we extract stellar masses and 
SFR values averaged over 10~Myr within twice the stellar half mass
radius for each MaNGIA galaxy. We then perform a linear fit to the 
Main Sequence, following methods described in Sec.~\ref{sec:DeltaMS},
arriving at an MS fit of 
\mbox{$\rm{MS}(M_\ast) = -8.00 + 0.78\, \rm{log}_{10}(M_\ast / [\rm M_\odot])$},
which is in close correspondence with the MS fit in the observed
MaNGA sample in Eq.~\ref{eq:msfr-manga}.

Fig.~\ref{fig:tng-stellar-fmr-lightweighted} shows a range in stellar 
metallicity offset to lower values w.r.t. Fig.~\ref{fig:mzr-TNG}. 
This way, the FMR is shifted closer towards the observations as 
seen in Fig.~\ref{fig:MaNGA_FMR}, however simulations remain concentrated
around values higher than those observed in MaNGA. Similar to Fig.~\ref{fig:mzr-TNG},
the simulated FMR is more compact than in the observations and 
different ranges in $\Delta_{\rm MS}$ show significant overlap 
within their errorbars. 

Although a direct comparison between 
Figs~\ref{fig:tng-stellar-fmr-lightweighted}~and~\ref{fig:MaNGA_FMR}
might suggest potential physical differences between the 
chemical composition of simulated and observed galaxy stellar contents,
one still needs to remain cautious about the challenges associated
with modelling the measurable electromagnetic signal emitted
by stellar populations in cosmological simulations. With simulations
currently lacking an explicit dust component and direct treatment
of the cold ISM, mock observation studies like those conducted
by \cite{Sarmiento2023} need to rely on a set of simplifying 
assumptions in order to estimate the impact of the ISM on the 
intrinsic stellar emission. These caveats, coupled with differences
in stellar parameter recovery among different continuum fitting
approaches, are not unlikely to yield quantitative differences 
in the inferred stellar FMR between observed and simulated 
galaxies. Therefore, having explored the light-weighted approach in the
mock TNG50-1 observation catalogue, we decide to focus on a qualitative
comparison of trends between MaNGA and IllustrisTNG in the main text.



\bibliographystyle{mnras}
\bibliography{Bib_TL}{}



\bsp	
\label{lastpage}
\end{document}